\documentclass[twocolumn]{aastex63}

\usepackage[utf8]{inputenc}
\usepackage{natbib}
\usepackage{amsmath}
\usepackage[style=american]{csquotes}
\usepackage{xspace}

\newcommand{\tess}{TESS\xspace}
\newcommand{\gaia}{\textit{Gaia}\xspace}
\newcommand{\rearth}{$R_{\oplus}$\xspace}
\newcommand{\mearth}{$M_{\oplus}$\xspace}
\newcommand{\msun}{M$_\odot$\xspace}
\newcommand{\rsun}{R$_\odot$\xspace}
\newcommand{\triceratops}{\texttt{TRICERATOPS}\xspace}
\newcommand{\vespa}{\texttt{vespa}\xspace}
\newcommand{\teff}{$T_{\mathrm{eff}}$}

\shorttitle{A Nearby Companion to WASP-132 b}
\shortauthors{Hord et al.}

\begin{document}

\title{The Discovery of a Planetary Companion Interior to Hot Jupiter WASP-132 b}

\author[0000-0001-5084-4269]{Benjamin J. Hord}
\affiliation{Department of Astronomy, University of Maryland, College Park, MD 20742, USA}
\affiliation{NASA Goddard Space Flight Center, 8800 Greenbelt Road, Greenbelt, MD 20771, USA}
\affiliation{GSFC Sellers Exoplanet Environments Collaboration}

\author[0000-0001-8020-7121]{Knicole D. Col\'{o}n}
\affiliation{NASA Goddard Space Flight Center, 8800 Greenbelt Road, Greenbelt, MD 20771, USA}
\affiliation{GSFC Sellers Exoplanet Environments Collaboration}

\author[0000-0002-2580-3614]{Travis A. Berger}
\altaffiliation{NASA Postdoctoral Program Fellow}
\affiliation{NASA Goddard Space Flight Center, 8800 Greenbelt Road, Greenbelt, MD 20771, USA}
\affiliation{GSFC Sellers Exoplanet Environments Collaboration}

\author[0000-0001-9786-1031]{Veselin Kostov}
\affiliation{NASA Goddard Space Flight Center, 8800 Greenbelt Road, Greenbelt, MD 20771, USA}
\affiliation{GSFC Sellers Exoplanet Environments Collaboration}

\author[0000-0003-2565-7909]{Michele L. Silverstein}
\altaffiliation{NASA Postdoctoral Program Fellow}
\affiliation{NASA Goddard Space Flight Center, 8800 Greenbelt Road, Greenbelt, MD 20771, USA}
\affiliation{GSFC Sellers Exoplanet Environments Collaboration}

\author[0000-0002-3481-9052]{Keivan G.\ Stassun}
\affiliation{Department of Physics and Astronomy, Vanderbilt University, Nashville, TN 37235, USA}

\author[0000-0001-6513-1659]{Jack J. Lissauer}
\affiliation{NASA Ames Research Center, Moffett Field, CA 94035, USA}

\author[0000-0001-6588-9574]{Karen A.\ Collins}
\affiliation{Center for Astrophysics \textbar \ Harvard \& Smithsonian, 60 Garden Street, Cambridge, MA 02138, USA}

\author[0000-0001-8227-1020]{Richard P. Schwarz}
\affiliation{Patashnick Voorheesville Observatory, Voorheesville, NY 12186, USA}

\author[0000-0003-3904-6754]{Ramotholo Sefako}
\affiliation{South African Astronomical Observatory, P.O. Box 9, Observatory, Cape Town 7935, South Africa}

\author[0000-0002-0619-7639]{Carl Ziegler}
\affiliation{Department of Physics, Engineering and Astronomy, Stephen F. Austin State University, 1936 North St, Nacogdoches, TX 75962, USA}

\author[0000-0001-7124-4094]{C\'{e}sar Brice\~{n}o}
\affiliation{Cerro Tololo Inter-American Observatory/NSF's NOIRLab, Casilla 603, La Serena, Chile}

\author[0000-0001-9380-6457]{Nicholas Law}
\affiliation{Department of Physics and Astronomy, The University of North Carolina at Chapel Hill, Chapel Hill, NC 27599-3255, USA}

\author[0000-0003-3654-1602]{Andrew W. Mann}
\affiliation{Department of Physics and Astronomy, The University of North Carolina at Chapel Hill, Chapel Hill, NC 27599-3255, USA}

\author[0000-0003-2058-6662]{George~R.~Ricker}
\affiliation{Department of Physics and Kavli Institute for Astrophysics and Space Research, Massachusetts Institute of Technology, Cambridge, MA 02139,
USA}

\author[0000-0001-9911-7388]{David~W.~Latham}
\affiliation{Center for Astrophysics \textbar \ Harvard \& Smithsonian, 60 Garden Street, Cambridge, MA 02138, USA}

\author[0000-0002-6892-6948]{S.~Seager}
\affiliation{Department of Physics and Kavli Institute for Astrophysics and Space Research, Massachusetts Institute of Technology, Cambridge, MA 02139,
USA}
\affiliation{Department of Earth, Atmospheric and Planetary Sciences, Massachusetts Institute of Technology, Cambridge, MA 02139, USA}
\affiliation{Department of Aeronautics and Astronautics, MIT, 77 Massachusetts Avenue, Cambridge, MA 02139, USA}

\author[0000-0002-4265-047X]{Joshua~N.~Winn}
\affiliation{Department of Astrophysical Sciences, Princeton University, 4 Ivy Lane, Princeton, NJ 08544, USA}

\author[0000-0002-4715-9460]{Jon~M.~Jenkins}
\affiliation{NASA Ames Research Center, Moffett Field, CA 94035, USA}

\author[0000-0002-0514-5538]{Luke G. Bouma}
\affiliation{Cahill Center for Astrophysics, California Institute of Technology, Pasadena, CA 91125, USA}

\author{Ben Falk}
\affiliation{Space Telescope Science Institute, 3700 San Martin Drive, Baltimore, MD, 21218, USA}

\author[0000-0002-5286-0251]{Guillermo~Torres}
\affiliation{Center for Astrophysics \textbar \ Harvard \& Smithsonian, 60 Garden Street, Cambridge, MA 02138, USA}

\author[0000-0002-6778-7552]{Joseph~D.~Twicken}
\affiliation{NASA Ames Research Center, Moffett Field, CA 94035, USA}
\affiliation{SETI Institute, Mountain View, CA 94043, USA}

\author[0000-0001-7246-5438]{Andrew~Vanderburg}
\affiliation{Department of Physics and Kavli Institute for Astrophysics and Space Research, Massachusetts Institute of Technology, Cambridge, MA 02139, USA}
\correspondingauthor{Benjamin J. Hord}
\email{benhord@astro.umd.edu}

\begin{abstract}
    Hot Jupiters are generally observed to lack close planetary companions, a trend that has been interpreted as evidence for high-eccentricity migration. We present the discovery and validation of WASP-132 c (TOI-822.02), a 1.85 $\pm$ 0.10 \rearth planet on a 1.01 day orbit interior to the hot Jupiter WASP-132 b. Transiting Exoplanet Survey Satellite (TESS) and ground-based follow-up observations, in conjunction with vetting and validation analysis, enable us to rule out common astrophysical false positives and validate the observed transit signal produced by WASP-132 c as a planet. Running the validation tools \vespa and \triceratops on this signal yield false positive probabilities of $9.02 \times 10^{-5}$ and 0.0107, respectively. Analysis of archival CORALIE radial velocity data leads to a 3$\sigma$ upper limit of 28.23 ms$^{-1}$ on the amplitude of any 1.01-day signal, corresponding to a 3$\sigma$ upper mass limit of 37.35 \mearth. Dynamical simulations reveal that the system is stable within the 3$\sigma$ uncertainties on planetary and orbital parameters for timescales of $\sim$100 Myr. The existence of a planetary companion near the hot Jupiter WASP-132 b makes the giant planet's formation and evolution via high-eccentricity migration highly unlikely. Being one of just a handful of nearby planetary companions to hot Jupiters, WASP-132 c carries with it significant implications for the formation of the system and hot Jupiters as a population.
\end{abstract}

\section{Introduction} \label{sec:intro}

Ever since the Nobel Prize-winning discovery of the first exoplanet around a Sun-like star \citep{mayor1995jupiter}, hot Jupiters have represented one of the greatest enigmas of exoplanet science. With radii $R_{\rm p} >$ 8 \rearth and orbital periods $P <$ 10 days \citep{winn2010hot, wang2015occurrence, garhart2020statistical}, hot Jupiters represent a class of planets with no analogue in our solar system. Traditional theories on planet formation are insufficient to explain the existence of giant gaseous planets so close to a host star \citep{lin1996orbital}. Therefore, new formation scenarios have been put forth (e.g. disk migration, planet-planet scattering, secular migration) to explain the existence of hot Jupiters, most of which involve an inward migration after initially forming beyond the ice line (e.g., \citealt{lin1996orbital, rasio1996dynamical}). However, none of these mechanisms alone can satisfy all observable constraints, leaving the primary pathways of hot Jupiter formation largely still uncertain \citep{dawson2014photoeccentric, dawson2018origins}.

One clue that may help distinguish between different formation pathways is that hot Jupiters are predominantly ``lonely", meaning they are the only planet in their system within a factor of 2 or 3 in orbital distance \citep[e.g.][]{steffen2012kepler, knutson2014friends, endl2014kepler, hord2021uniform}, although they may have more distant companions, particularly giant companions \citep{schlaufman2016occurrence}. This lack of nearby companions is expected from a high-eccentricity migration formation pathway -  a scenario in which the hot Jupiter migrates inwards from beyond the ice line via some form of gravitational perturbations that put it on an eccentric orbit that eventually circularizes much closer to the host star \citep{rasio1996dynamical}. This high-eccentricity migration results in the scattering and possible ejection of other planets in the system as the hot Jupiter's eccentric orbit sweeps through the inner parts of the stellar system \citep{mustill2015destruction}.

Of the $\sim$500 hot Jupiters currently confirmed, only three have proven to be exceptions to this lonely trend so far. The systems WASP-47 \citep{becker2015wasp}, Kepler-730 \citep{canas2019kepler}, and TOI-1130 \citep{huang2020tess} all host a hot Jupiter with at least one nearby companion planet, making the high-eccentricity migration scenario for these planetary systems unlikely\footnote{We note WASP-148b is a hot Jupiter with an outer, massive companion at just within 3 times the orbital distance of the hot Jupiter. We choose not to include this system in our discussion of hot Jupiters with nearby companions because of its very different architecture from the other systems  (which have smaller, closer companions) and because the mass and orbital distance of the companion planet puts it on the borderline of the approximate definition of  hot Jupiters with close planetary companions.}. It is more likely that these planetary systems formed via disk migration, where the protoplanets migrate inwards all together within the disk, potentially preserving planets near the hot Jupiter \citep{lin1996orbital, lee2002dynamics, raymond2006exotic}. These systems serve as rare opportunities to dynamically constrain the formation of hot Jupiters and also potentially serve as a bridge to the slightly cooler population of warm Jupiters ($10<P<100$ d), which are often joined by smaller companion planets \citep{huang2016warm}. Hot Jupiters with nearby planets may be key in understanding the connection of formation pathways to the observed hot Jupiter population.

The Transiting Exoplanet Survey Satellite (\tess, \citealt{ricker2015transiting}) is well suited to the discovery of hot Jupiters and potential nearby companions, as its almost-all-sky coverage is expected to observe nearly every known hot Jupiter system and discover hundreds or thousands more \citep{sullivan2015transiting, barclay2018revised}. This is particularly important since the hot Jupiter sample is currently heterogeneous and incomplete \citep{yee2021complete}. In addition, TESS has the photometric precision to identify smaller planets down to $\sim$0.7 \rearth \citep[e.g.][]{kostov2019b, gilbert2020first, silverstein2022lhs}. 

Here we present the \tess discovery of TOI-822.02 --- henceforth referred to as WASP-132 c --- a small planet associated with hot Jupiter WASP-132 b first discovered by \cite{2017MNRAS.465.3693H}. The new planet WASP-132 c is on a 1.01 d orbit interior to the 7.13 d orbit of the hot Jupiter WASP-132 b. This makes the WASP-132 system the fourth such system containing a hot Jupiter with a nearby small planetary companion, widening the sample of this rare subclass of hot Jupiters and further opening the possibility for comparative planetology both within the hot Jupiter system and between hot and warm Jupiter systems. 

Section \ref{sec:search_vetting} details the discovery of the WASP-132 c signal in the \tess photometric data as well as the initial vetting efforts. Section \ref{sec:stellar_params} presents our refinement of the stellar parameters of the host star using a series of independent models. Section \ref{sec:modeling} describes our methodology for modeling the full photometric light curve to obtain precise planetary and orbital parameters for WASP-132 c as well as the confirmed hot Jupiter WASP-132 b. Section \ref{sec:validation} presents the validation of WASP-132 c as a planet based on ground-based follow-up observations and statistical validation software. Section \ref{sec:dynamics} details the dynamical simulations modeling a two-planet WASP-132 system to probe the long-term stability of the system. Section \ref{sec:discussion} discusses the implications of the discovery of such a system in terms of hot Jupiter formation and the larger hot Jupiter sample.

\section{Signal Search and Vetting} \label{sec:search_vetting}

WASP-132 (TOI-822, TIC 127530399) was observed by \tess in Sector 11 from UT April 23 to May 20, 2019 (23.96 d) in CCD 2 of Camera 1 and in Sector 38 from UT April 29 to May 26, 2021 (26.34 d) in CCD 1 of Camera 1. Data for WASP-132 were collected at 2 minute cadence in Sectors 11 and 38 and at 20 second cadence in Sector 38. The star was prioritized for high-cadence measurements as part of the Cycle 1 Guest Investigator Programs G011112, G011183, and G011132 and Cycle 3 Guest Investigator Programs G03278, G03181, and G03106.\footnote{Details of approved \tess Guest Investigator Programs are available from \url{https://heasarc.gsfc.nasa.gov/docs/tess/approved-programs.html}}

\begin{figure*}
    \centering
    \includegraphics[width=\textwidth]{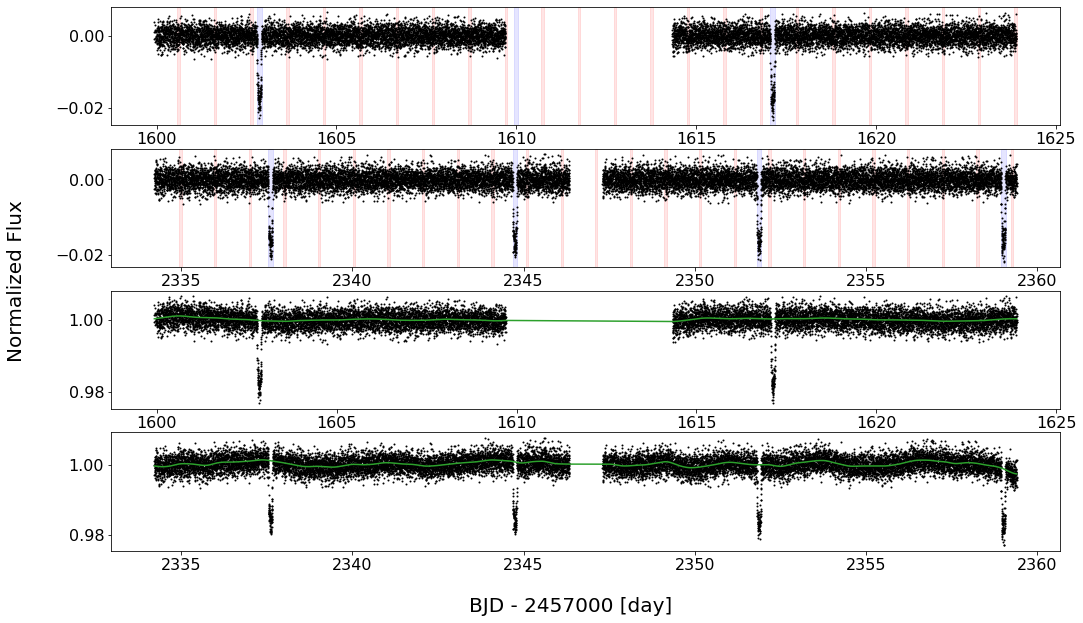}
    \caption{\tess PDC$\_$SAP light curve of the WASP-132 system with the in-transit times of hot Jupiter WASP-132 b (blue) and WASP-132 (red) highlighted. Both Sector 11 (top) and Sector 38 (second from top) are shown. The data is detrended with a Gaussian Process according to the method outlined in Section \ref{sec:modeling}. These Gaussian Process noise models are shown in green for Sector 11 (second from bottom) and Sector 38 (bottom) overlaid on the PDC$\_$SAP light curves to show how they capture the variability in the light curve.}
    \label{fig:tess_lightcurve}
\end{figure*}

The \tess Science Processing Operation Center (SPOC) pipeline \citep{Jenkins2016} processed the short cadence pixel data and generated the target pixel files (TPFs) and light curves cleaned of instrumental systematics. The Transiting Planet Search module (TPS, \citealp{jenkins2002impact, 2010SPIE.7740E..0DJ, jenkins2020kepler}) of the SPOC pipeline searched the generated 2-minute light curves for periodic, transit-like signals for each \tess sector independently and jointly as a single light curve. TPS recovered the previously confirmed hot Jupiter WASP-132 b as well as a new signal at 1.01153 d with a signal-to-noise ratio (SNR) of 10.6 in the combined data from the two sectors. This signal's depth corresponds to a planet with a radius of 2.33\rearth when using the stellar radius value for this target contained in the \tess Input Catalog (TICv8.2, \citealp{stassun2018tess, stassun2019revised}) to calculate the potential planet radius. The transits of the hot Jupiter and WASP-132 c in the \tess data can be seen in Figure \ref{fig:tess_lightcurve}.

To provide an independent recovery of this periodic signal, we searched the available WASP-132 \tess light curves with the Transit Least Squares (TLS) search algorithm \citep{2019A&A...623A..39H}. TLS utilizes analytical transit shapes, making it more sensitive to planet transits than the conventional Box Least Squares (BLS, \citealp{kovacs2002box}) search method and more finely tuned to detect small, short-period planets such as WASP-132 c.

Our transit search with TLS made use of the systematics-corrected Presearch Data Conditioning Simple Aperture Photometry (PDC$\_$SAP) \tess light curves generated by the TESS SPOC pipeline \citep{smith2012kepler, stumpe2012kepler, stumpe2014multiscale} at the 2 minute and 20 second cadence for \tess Sectors 11 and 38, respectively\footnote{We elected to use the shortest cadence available for each sector to capture the shape of the transit as accurately as possible.}. We used the \texttt{lightkurve} Python package \citep{2018ascl.soft12013L} to download the data from the Mikulski Archive for Space Telescopes (MAST). The light curve exhibited small amplitude stellar variability and was detrended using \texttt{lightkurve}'s built in \texttt{flatten} method. A window length of $>$0.5 days was chosen as it was large enough to preserve transit signals (of duration on the order of $\lesssim$3 hours) while small enough to remove the slight stellar variability present in the light curve.

Using TLS on the clean light curve, we recovered the known signal of hot Jupiter WASP-132 b as well as a signal with a period of 1.0119 $\pm$ 0.0032 d with a false alarm probability (FAP) $<10^{-4}$, which is well below the threshold of what \cite{2019A&A...623A..39H} states is a significant detection above white noise. The period, depth, and mid-transit time of this recovered signal are consistent with the values reported by the SPOC pipeline and listed on the Exoplanet Follow-up Observing Program-TESS (ExoFOP-TESS, \citealt{ExoFOP2019-wy}) website\footnote{\url{https://exofop.ipac.caltech.edu/tess/}}. This recovery with TLS served as an independent check to ensure that the signal was not a pipeline-specific detection.

We performed multiple initial checks of the signal and \tess light curves for astrophysical false-positive scenarios that can mimic exoplanet transits. The Data Validation module (DV, \citealp{Twicken:DVdiagnostics2018, Li:DVmodelFit2019}) of the SPOC pipeline performs a suite of diagnostic vetting tests to investigate the likelihood of many of these false-positive scenarios. These tests include a depth test of the odd and even transits, a statistical bootstrap test that accounts for the non-white nature of the light curve to estimate the probability of a false alarm from random noise fluctuations, a ghost diagnostic test to compare the detection statistic of the optimal aperture against that of a halo with a 1 pixel buffer around the optimal aperture, and a difference image centroid test. WASP-132 c passed all of these diagnostic vetting tests. Additionally, all TICv8.2 objects other than the target star were statistically excluded as sources of the 1.01 d transit signal since the difference centroid offset tests located the source of the transit signal to within 1 $\pm$ 3 arcsec of the target position. The Threshold Crossing Event (TCE) was promoted to TESS Object of Interest \citep[TOI;][]{guerrero2021tess} status and designated TOI-822.02 by the TESS Science Office based on the clean model fit and diagnostic test results in the SPOC data validation report.

In addition to the vetting checks performed by the SPOC pipeline, we used \texttt{DAVE} (Discovery and Vetting of Exoplanets, \citealp{kostov2019a}) to further check for astrophysical false-positive scenarios. \texttt{DAVE} is an automated vetting pipeline built upon many of the tools developed for vetting planets in \textit{Kepler} data (e.g. RoboVetter, \citealp{coughlin14}). It has been used extensively in vetting planets in \textit{K2} (\citealp{hedges19, 2021MNRAS.508..195D}) and \tess (\citealp{kostov2019b, crossfield19, 2020AJ....160..116G}) data as well. \texttt{DAVE} performs two sets of vetting tests. The first are light curve-based vetting tests searching for odd/even transit differences, secondary eclipses, and light curve modulations that could introduce transit-like signals. The second set of tests are image-based that check the photocenter motion on the \tess image during transit.

Unfortunately, due to the weak signal resulting from the shallow transit depth and relatively dim ($Tmag$=11.11) stellar host, many of the results from \texttt{DAVE} were inconclusive but still showed no significant indication of an astrophysical false-positive scenario. As such, we determined that further, more detailed modeling of the transits was warranted.

\section{Refinement of Host Star Parameters} \label{sec:stellar_params}

To perform a comprehensive analysis of the transit and system light curve, it was necessary to refine the stellar parameters of the host star, WASP-132. The TICv8.2 reports key stellar parameters determined via independent analysis, but there are also stellar parameters reported by the WASP Collaboration in the initial discovery and confirmation of the hot Jupiter WASP-132 b \citep{2017MNRAS.465.3693H}. We performed our own independent analysis given available data in order to determine the best values for the stellar parameters to use when modeling the \tess data. The results of each independent analysis method are contained in Table \ref{tab:params_compare} for comparison. We find most stellar parameter values from each analysis method are consistent with each of the others as well as with both those reported by \cite{2017MNRAS.465.3693H} and the TICv8.2 \citep{stassun2018tess, stassun2019revised}. The exception is a slight difference between the bolometric flux $F_{\rm bol}$ reported by the two SED analyses. The adopted stellar parameter values used in the remaining validation and analysis of WASP-132 c are based on the isochrone analysis described below and are contained in Table \ref{tab:stellar_params}.

We also note that we see a low-amplitude 8 d periodic variation upon visual inspection of the Sector 38 light curve that does not match up with the 33 day stellar rotation period stated in \cite{2017MNRAS.465.3693H}. If the 8 d variability were to represent the rotation period of the star, this would imply a $v$ sin $i$ of $\sim$ 5 km s$^{-1}$, using the equation $v$ sin $i$ = (2$\pi$R)/P and assuming the star is viewed edge-on. This is well outside of the confidence interval of the measured value $v$ sin $i$ = 0.9 $\pm$ 0.8 km s$^{-1}$, suggesting that this 8 d variability is not the stellar rotation period, although it may represent one of the harmonics of the true period if it is intrinsic to the star. However, extracted light curves using different apertures do not contain this $\sim$8 day variation, suggesting that this shorter-scale variability may not be inherent to the WASP-132 system. Regardless of the origin of this additional variability, our conclusions remain the same regarding the planetary nature of the 1.01 d transit signal.

Overall, WASP-132 does not show significant signs of activity. There is the possible 8 d variability and reported 33 d rotation period, activity that occurs on much longer timescales than the orbital period of WASP-132 c. We also found no evidence of flares or star spot crossings in either the space- or ground-based data. In the modeling of the system's light curves (see Section \ref{sec:modeling}), a Gaussian Process was used to capture any variability in conjunction with transit models for each planet. Thus, photometric variability was modeled out while not diluting the transit signals (see Figure \ref{fig:tess_lightcurve}), in order to precisely measure the planet and orbital parameters.

\begin{deluxetable*}{cccccc}
\tablecaption{Stellar parameters obtained using each of the methods outlined in Section \ref{sec:stellar_params}. Values from \cite{2017MNRAS.465.3693H}, which discovered WASP-132 b, and from the TICv8.2 \citep{stassun2018tess, stassun2019revised} are included for comparison. The final adopted parameters for this analysis are contained in Table \ref{tab:stellar_params}. All uncertainties reported are the 1$\sigma$ value.}
\label{tab:params_compare}
\tablehead{ \colhead{Parameter} & \colhead{TICv8.2} & \colhead{\cite{2017MNRAS.465.3693H}} & \colhead{Isochrone Analysis} & \colhead{KGS SED Analysis} & \colhead{MLS SED Analysis} }
\startdata
$T_{\rm eff}$ (K) & 4742 $\pm$ 129 & 4750 $\pm$ 100 & $4714\substack{+87 \\ -88}$ & 4750 $\pm$ 75 & 4753 $\pm$ 80 \\
{[Fe/H]} & --- & 0.22 $\pm$ 0.13 & 0.18 $\pm$ 0.12 & 0.0 $\pm$ 0.5 & --- \\
$M_{*}$ (\msun) & 0.760 $\pm$ 0.089 & 0.80 $\pm$ 0.04 & 0.782 $\pm$ 0.034 & 0.80 $\pm$ 0.05 & --- \\
$R_{*}$ (\rsun) & 0.790 $\pm$ 0.057 & 0.74 $\pm$ 0.02 & $0.753\substack{+0.028 \\ -0.026}$ & 0.752 $\pm$ 0.024 & 0.767 $\pm$ 0.026 \\
$L_{*}$ ($L_{\odot}$) & 0.284 $\pm$ 0.012 & --- & $0.253\substack{+0.032 \\ -0.028}$ & --- & 0.271 $\pm$  0.007\\
log(g) & 4.524 $\pm$ 0.094 & 4.6 $\pm$ 0.1 & $4.576\substack{+0.028 \\ -0.036}$ & --- & --- \\
$\rho_{*}$ (g cm$^{-3}$) & 2.17 $\pm$ 0.59 & $2.82\substack{+0.10 \\ -0.20}$ & $1.81\substack{+0.18 \\ -0.19}$ & --- & --- \\
Age (Gyr) & --- & $\gtrsim$0.5 & $7.055\substack{+7.114 \\ -5.000}$ & 3.2 $\pm$ 0.5 & --- \\
Distance (pc) & 122.91 $\pm$ 0.57 & 120 $\pm$ 20 & 126 $\pm$ 5 & --- & --- \\
$F_{\rm bol}$ (erg s$^{-1}$ cm$^{-2}$ $\times$10$^{-10}$) & --- & --- & --- & 5.442 $\pm$ 0.063 & 5.69 $\pm$ 0.14 \\
\enddata
\end{deluxetable*}

\begin{deluxetable}{llr}
\tablecaption{Adopted stellar parameters for WASP-132.}
\label{tab:stellar_params}
\tablehead{ \colhead{Parameter} & \colhead{Value} & \colhead{Source} }
\startdata
\multicolumn{3}{c}{\em Identifying Information} \\
Name & WASP-132 & \\
TIC ID & 127530399 & TICv8.2 \\
TOI ID & TOI-822 & \cite{guerrero2021tess} \\
Alt. Name & UCAC4 220-083803 & \\
 & & \\
\multicolumn{3}{c}{\em Astrometric Properties} \\
$\alpha$ R.A. (hh:mm:ss) & 14:30:26.21 (J2015.5) & \gaia EDR3 \\
$\delta$ Dec. (dd:mm:ss) & -46:09:34.29 (J2015.5) & \gaia EDR3 \\
$\mu_{\alpha}$ (mas yr$^{-1}$) & 12.255 $\pm$ 0.020 & \gaia EDR3 \\
$\mu_{\delta}$ (mas yr$^{-1}$) & -73.169 $\pm$ 0.022 & \gaia EDR3 \\
Distance (pc) & 123.57 $\pm$ 0.29 & \gaia EDR3 \\
\multicolumn{3}{c}{\em Stellar Properties} \\
Spectral Type & K4 & \cite{2017MNRAS.465.3693H} \\
$T_{\rm eff}$ (K) & $4714\substack{+87 \\ -88}$ & This Work \\
{[Fe/H]} & 0.18 $\pm$ 0.12 & This Work \\
$M_{*}$ (\msun) & 0.782 $\pm$ 0.034 & This Work \\
$R_{*}$ (\rsun) & $0.753\substack{+0.028 \\ -0.026}$ & This Work \\
$L_{*}$ ($L_{\odot}$) & $0.253\substack{+0.032 \\ -0.028}$ & This Work \\
log(g) & $4.576\substack{+0.028 \\ -0.036}$ & This Work \\
$\rho_{*}$ (g cm$^{-3}$) & $1.81\substack{+0.18 \\ -0.19}$ & This Work \\
Rotation period (d) & $\sim$33 & \cite{2017MNRAS.465.3693H} \\
$v$ sin $i$ (km s$^{-1}$) & 0.9 $\pm$ 0.8 & \cite{2017MNRAS.465.3693H} \\
Age (Gyr) & 3.2 $\pm$ 0.5 & This Work \\
\multicolumn{3}{c}{\em Photometric Properties} \\
$B$ (mag) & 13.142 $\pm$ 0.011 & APASS DR9 \\
$V$ (mag) & 11.938 $\pm$ 0.046 & APASS DR9 \\
$G_{\rm G}$ (mag) & 11.7467 $\pm$ 0.0002 & \gaia EDR3 \\
$G_{\rm BP}$ (mag) & 12.3000 $\pm$ 0.0007 & \gaia EDR3\\
$G_{\rm RP}$ (mag) & 11.0487 $\pm$ 0.0004 & \gaia EDR3 \\
$T$ (mag) & 11.111 $\pm$ 0.006 & TICv8.2 \\
$J$ (mag) & 10.257 $\pm$ 0.026 & 2MASS \\
$H$ (mag) & 9.745 $\pm$ 0.023 & 2MASS \\
$K_{\rm s}$ (mag) & 9.674 $\pm$ 0.024 & 2MASS \\
$W_{1}$ (mag) & 9.557 $\pm$ 0.022 & AllWISE \\
$W_{2}$ (mag) & 9.638 $\pm$ 0.020 & AllWISE \\
$W_{3}$ (mag) & 9.575 $\pm$ 0.040 & AllWISE \\
$W_{4}$ (mag) & 8.281$^1$ & AllWISE \\
\enddata
\tablenotetext{}{\gaia EDR3 - \cite{prusti2016gaia, brown2021gaia}, TICV8.2 - \cite{stassun2019revised}, APASS DR9 - \cite{Henden2016}, 2MASS - \cite{Skrutskie2006}, AllWISE - \cite{Cutri2013}. $^1$Only a limit is reported for $W_{4}$ since the signal-to-noise ratio was too low for a confident detection.}
\end{deluxetable}

\subsection{Isochrone Analysis}

We performed an isochrone-based analysis for WASP-132 using \texttt{isoclassify} \citep{huber2017,berger2020}, which produces fundamental stellar parameters from a combination of input observables. We used spectroscopic \teff\ and metallicity from the discovery paper \citep{2017MNRAS.465.3693H}, $Gaia$ Data Release 2 \citep[DR2]{, prusti2016gaia, gaia2018, bailerjones2018} parallax and coordinates, and the Two Micron All-Sky Survey \citep[{\it 2MASS}]{Skrutskie2006} $K_{\rm s}$ magnitude as inputs. We also used the \texttt{allsky} extinction map detailed in \citet{bovy2016} to estimate the photometric extinction based on the coordinates and distance inferred from the parallax. The best-fit values and their uncertainties are compiled in Table \ref{tab:params_compare}, and we estimate extinction to be $A_V$ = 0.097 $\pm$ 0.024 mag, which is consistent with the $A_V$ = 0.093 $\pm$ 0.031 mag reported in the TICv8.2.

\subsection{KGS SED Analysis} \label{ssec:KGS_SED_analysis}

\begin{figure}[!ht]
    \centering
    \includegraphics[width=0.48\textwidth,trim=90 70 90 90,clip]{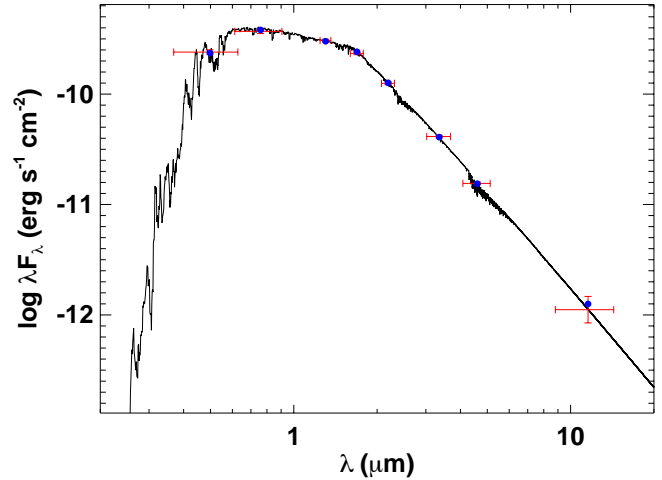}
    \caption{Spectral energy distribution of WASP-132. Red symbols represent the observed photometric measurements outlined in Section \ref{ssec:KGS_SED_analysis}, where the horizontal bars represent the effective width of the passband. Blue symbols are the model fluxes from the best-fit NextGen atmosphere model (black).}
    \label{fig:sed}
\end{figure}

As an independent determination of the basic stellar parameters, K.G. Stassun (KGS) performed an analysis of the broadband spectral energy distribution (SED) of the star together with the {\it Gaia\/} Early Data Release 3 \citep[EDR3]{prusti2016gaia, brown2021gaia} parallax \citep[with no systematic offset applied; see, e.g.,][]{StassunTorres:2021}, in order to determine an empirical measurement of the stellar radius, following the procedures described in \citet{Stassun:2016,Stassun:2017,Stassun:2018}. We pulled the the $JHK_S$ magnitudes from {\it 2MASS}, the W1--W3 magnitudes from the Wide-field Infrared Survey Explorer \citep[{\it WISE}]{wright2010wide, Cutri2013}, and the $G_{\rm BP} G_{\rm RP}$ magnitudes from {\it Gaia}. Together, the available photometry spans the full stellar SED over the wavelength range 0.4--10~$\mu$m (see Figure~\ref{fig:sed}).  

We performed a fit using NextGen stellar atmosphere models, with the free parameters being the effective temperature ($T_{\rm eff}$) and metallicity ([Fe/H]), as well as the extinction $A_V$, which we fixed at zero due to the proximity of the system to Earth. The resulting fit (Figure~\ref{fig:sed}) has a best-fit $T_{\rm eff} = 4750 \pm 75$~K, [Fe/H] = $0.0 \pm 0.5$, with a reduced $\chi^2$ of 0.8. Integrating the (unreddened) model SED gives the bolometric flux at Earth, $F_{\rm bol} = 5.442 \pm 0.063 \times 10^{-10}$ erg~s$^{-1}$~cm$^{-2}$. Taking the $F_{\rm bol}$ and $T_{\rm eff}$ together with the {\it Gaia\/} parallax gives the stellar radius, $R_\star = 0.752 \pm 0.024$~R$_\odot$. In addition, we can estimate the stellar mass from the empirical relations of \citet{Torres:2010}, giving $M_\star = 0.80 \pm 0.05$~M$_\odot$. Finally, the reported stellar rotation period of $\sim$33 d implies an age of $\tau_\star = 3.2 \pm 0.5$~Gyr via the empirical gyrochronology relations of \citet{Mamajek:2008}. 

\subsection{MLS SED Analysis}

As an additional check on our stellar effective temperature, luminosity, and radius results, M. L. Silverstein (MLS) led a second SED analysis following methodology similar to \cite{dieterich2014} (Silverstein et al., in preparation). Spanning optical to mid-infrared wavelengths, we extract $JHK_S$ magnitudes from {\it 2MASS} and W1--W3 magnitudes from {\it WISE} as in the previous subsection. We differ in our adoption of the {\it Gaia\/}~DR2 parallax and of $VRI$ photometry converted from {\it Gaia\/}~DR2 $G_{\rm G} G_{\rm BP} G_{\rm RP}$. We compare nine different color combinations to the BT-Settl 2011 photospheric models \citep{allard2012} to derive a best-fitting $T_{\rm eff} = 4753 \pm 80$~K, assuming [Fe/H] = 0. Next we iteratively scale the resulting best-match model using a polynomial function until model and observed photometry match to within the error bars. We then integrate the final spectrum and apply a bolometric correction to determine $F_{\rm bol} = 5.691 \pm 0.137 \times 10^{-10}$ erg~s$^{-1}$~cm$^{-2}$, and we scale by the parallax to derive $L_* = 0.271 \pm 0.007~L_\odot$. A radius of $R_* = 0.767 \pm 0.026~R_\odot$ is then calculated using the Stefan-Boltzmann Law. These results are listed in Table~\ref{tab:params_compare} and match those from the other independent methods described in this paper. 

\section{Modeling the Physical Properties of WASP-132 c} \label{sec:modeling}

While TLS is useful at detecting signals, the period grid that it searches is not very fine by default. Combined with the refined stellar parameters discussed in Section \ref{sec:stellar_params}, it is possible to model the light curve in a more detailed fashion than the initial transit search to find the maximum likelihood values for the planet properties in the system. To perform this detailed modeling, we used the software \texttt{exoplanet} \citep{exoplanet:exoplanet}. \texttt{exoplanet} is a toolkit for probabilistic modeling of transit and radial velocity observations of exoplanets using PyMC3. This is a powerful and flexible program that can be used to build high-performance transit models and then sample them through Markov Chain Monte Carlo (MCMC) simulations to provide precise transit and orbital parameters.

We utilized the same PDC$\_$SAP light curves used in the transit search with TLS with one difference. For the modeling with \texttt{exoplanet}, we did not apply any initial detrending that could possibly alter the transit signals but instead included a Gaussian Process (GP) in the model. Our model had three elements: two planet components with Keplerian orbits and limb-darkened transits (one for each potential planet in the system) and a GP component that modeled residual stellar variability. The planet models were computed using STARRY \citep{luger2019starry} and the GP was computed using \texttt{celerite} \citep{foreman2017fast, 2018RNAAS...2...31F}. The GP component models the residual stellar variability in the light curve and describes a stochastically-driven, damped harmonic oscillator with two hyper-parameters, ln($S_{0}$) and ln($\omega_{0}$), which represent the undamped angular frequency of a simple harmonic oscillator and the power at $\omega$ = 0, respectively. We fixed the quality factor Q of the simple harmonic oscillator to 1 / $\sqrt{2}$ and put wide Gaussian priors on ln($S_{0}$) and ln($\omega_{0}$), setting their means to the natural log of the standard deviation of the flux and natural log of one tenth of a cycle, respectively, with both of their standard deviations set to 10. This form of GP has the advantage of being able to model a wide range of low frequency astrophysical and instrumental signals without requiring a physical model for the observed variability. We also included a white noise term in the model which is parameterized by the natural log of the standard deviation of the flux with a prior identical to that of ln($S_{0}$). The GP parameters of the two \tess sectors were modeled separately since the sectors may have different noise parameters, especially since the data were taken at different cadences.

The planet model was parameterized with a two term limb-darkening component and the stellar radius and mass. The individual planets were parameterized in terms of the natural log of orbital period, mid-transit time, transit depth, impact parameter, eccentricity, and periastron angle at time of transit. For our priors on the stellar parameter components of the model, we used the mean and standard deviation values of our analysis discussed in Section \ref{sec:stellar_params} and displayed in Table \ref{tab:stellar_params}. We followed \cite{kipping2013efficient} for the parameterization of the limb-darkening. We used the SPOC values listed on ExoFOP-TESS for the means and standard deviations of Gaussian priors on the natural log of the orbital period, mid-transit time, and transit depth for the two planets. We imposed a uniform prior on the impact parameter bounded between 0 and 1. For the eccentricity prior, we used a Beta prior with $\alpha$ = 0.867 and $\beta$ = 3.03 as suggested by \cite{kipping2013parametrizing}. The eccentricity was bounded between zero and one and sampled as $e$cos($\omega$). The periastron angle at transit was sampled in vector space to avoid the sampler seeing discontinuities. We sampled the posterior distribution of the model parameters using the No U-turn Sampler (NUTS, \citealp{hoffman2014no}), which is a form of Hamiltonian Monte Carlo, as implemented by \texttt{PyMC3} \citep{salvatier2016probabilistic}. We ran 3 simultaneous chains, each with 2000 tuning steps and 2500 draws in the final sample.

\begin{deluxetable}{lc}
\tablecaption{Planet and orbital parameters for WASP-132 b and c calculated by modeling the \tess photometric data with \texttt{exoplanet}. Errors are reported at the 1$\sigma$ level. Noise parameters are also included.}
\label{tab:planet_params}
\tablehead{ \colhead{Parameter} & \colhead{Value} }
\startdata
\multicolumn{2}{c}{\em Model Parameters} \\
\textbf{Star} & \\
Limb darkening $u_{1}$ & 0.43 $\pm$ 0.11 \\
Limb darkening $u_{2}$ & 0.17 $\pm$ 0.24 \\
Radius [\rsun] & 0.754 $\pm$ 0.024 \\
Mass [\msun] & 0.781 $\pm$ 0.033 \\
ln($\rho_{\rm GP}$) & 1.03 $\pm$ 0.15 \\
ln($\sigma_{\rm GP, S11}$) & -7.96 $\pm$ 0.16 \\
ln($\sigma_{\rm GP, S38}$) & -7.11 $\pm$ 0.16 \\
& \\
\textbf{WASP-132 c} & \\
$T_{0}$ (BJD - 2457000) & 1597.5762 $\pm$ 0.0024 \\
ln(Period) [days] & 0.011 $\pm$ 4.69e-6 \\
Impact parameter & $0.28\substack{+0.24 \\ -0.19}$ \\
ln(Transit Depth) & -7.437 $\pm$ 0.068 \\
eccentricity & $0.13\substack{+0.20 \\ -0.09}$ \\
$\omega$ [radians] & $-0.93\substack{+1.83 \\ -1.65}$ \\
& \\
\textbf{WASP-132 b} & \\
$T_{0}$ (BJD - 2457000) & 2337.6080 $\pm$ 0.0002 \\
ln(Period) [days] & 1.96 $\pm$ 5.49e-7 \\
Impact parameter & 0.16 $\pm$ 0.11 \\
ln(Transit Depth) & -4.026 $\pm$ 0.01 \\
eccentricity & $0.07\substack{+0.15 \\ -0.05}$ \\
$\omega$ [radians] & $-0.08\substack{+2.55 \\ -2.67}$ \\
\hline
\multicolumn{2}{c}{\em Derived Parameters} \\
\textbf{WASP-132 c} & \\
Period [days] & 1.011534 $\pm$ 0.000005 \\
$R_{\rm p}/R_{*}$ & 0.023 $\pm$ 0.001 \\
Radius [\rearth] & 1.85 $\pm$ 0.10 \\
Radius [$R_{\rm J}$] & 0.165 $\pm$ 0.009 \\
$a/R_{*}$ & 5.17 $\pm$ 0.18 \\
$a$ [AU] & 0.0182 $\pm$ 0.0003 \\
Inclination [deg] & $86.64\substack{+1.12 \\ -3.52}$ \\
Duration [hours] & $1.47\substack{+0.14 \\ -0.22}$ \\
& \\
\textbf{WASP-132 b} & \\
Period [days] & 7.133514 $\pm$ 0.000004 \\
$R_{\rm p}/R_{*}$ & 0.122 $\pm$ 0.006 \\
Radius [\rearth] & 10.05 $\pm$ 0.34 \\
Radius [$R_{\rm J}$] & 0.897 $\pm$ 0.030 \\
$a/R_{*}$ & 19.03 $\pm$ 0.66 \\
$a$ [AU] & 0.067 $\pm$ 0.001 \\
Inclination [deg] & $89.51\substack{+0.14 \\ -0.49}$ \\
Duration [hours] & $3.18\substack{+0.18 \\ -0.21}$ \\
\enddata
\end{deluxetable}

\begin{figure}
    \centering
    \includegraphics[width=0.48\textwidth]{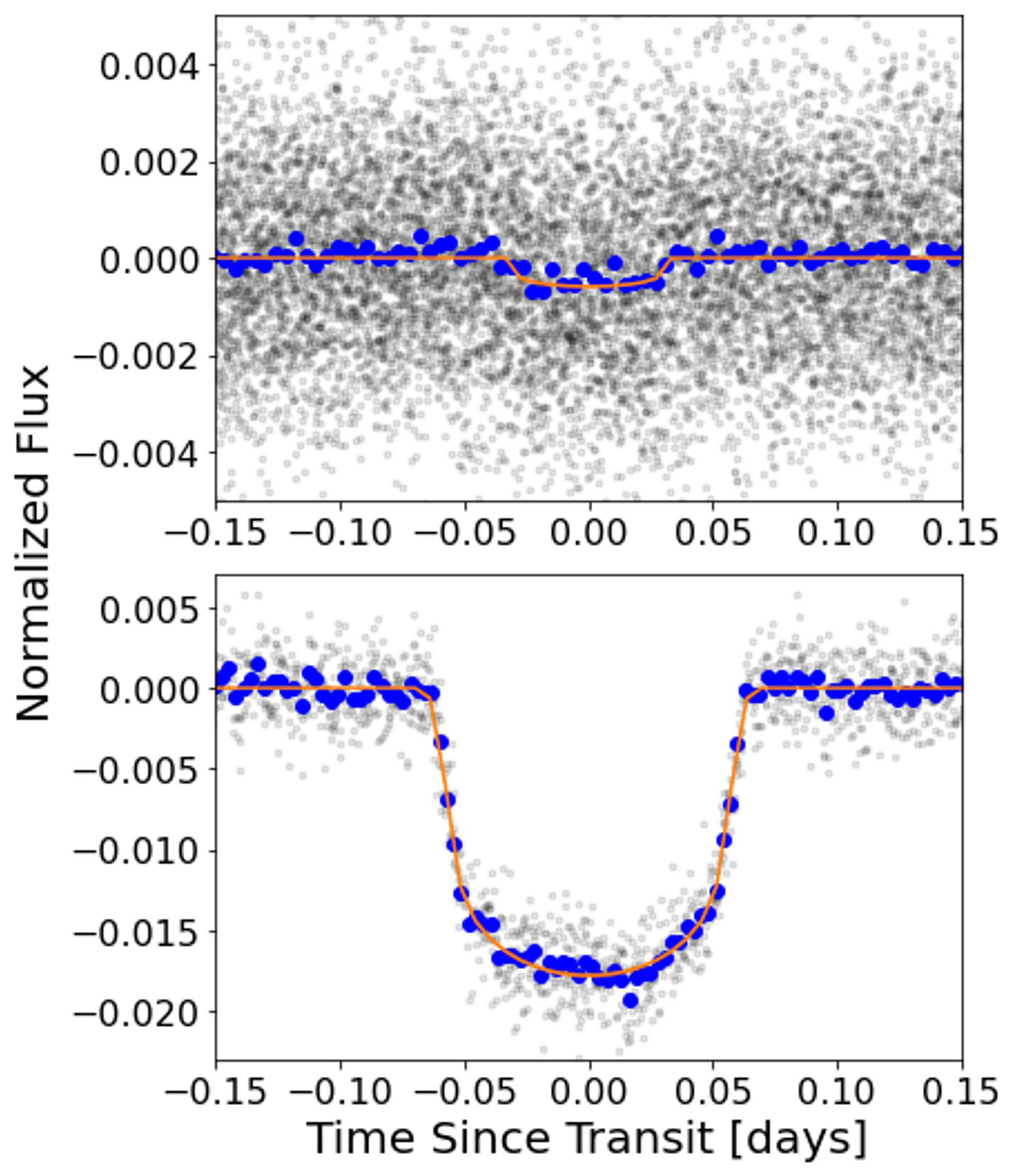}
    \caption{\tess data (light gray) for WASP-132 c (top) and WASP-132 b (bottom) phase folded to the best-fit period and mid-transit time with the \texttt{exoplanet} model overlaid. The process of fitting the transit model is described in Section \ref{sec:modeling}. The blue points are the phase-folded photometric data binned for clarity.}
    \label{fig:folded_transit}
\end{figure}

Initially, since individual \tess sectors often have different noise properties, we modeled both Sectors 11 and 38 independently from one another using the model described above. However, the resulting posterior distributions were equivalent within 1$\sigma$ errors, so we decided to combine both sectors of \tess data into a single light curve and use the same model. Since the two light curves are separated by $>$1 year, modeling both together as a single light curve increases the time baseline with which to model the orbit of the system, allowing for a better constrained orbital period than any individual sector or two sectors back-to-back in time. We binned the Sector 38 \tess data from 20 second cadence to 2 minute cadence to match with the \tess Sector 11 data in order to create a uniform data set with an increased photometric precision in Sector 38. No correction for contamination from nearby sources of constant brightness was included since follow up observations by SOAR and LCOGT cleared the nearby field of contaminating sources of this nature (see Section \ref{ssec:follow-up}).

The median values and 1$\sigma$ errors for the best-fit transit model are contained in Table \ref{tab:planet_params} and the folded light curve with the best-fit and 1$\sigma$ bounds of the transit model are shown in Figure \ref{fig:folded_transit}.

\section{Validation of WASP-132 c} \label{sec:validation}

While the \tess pipeline and \texttt{DAVE} perform vetting analysis against possible false alarm and false positive scenarios, they alone are not rigorous enough to validate the planetary nature of WASP-132 c. We therefore investigated this signal using both observational constraints (Section \ref{ssec:follow-up}) as well as publicly-available statistical software packages \vespa and \triceratops (Section \ref{ssec:software_validation}).

\subsection{Follow-up Observations} \label{ssec:follow-up}

In order to better constrain the false positive probability of a planet candidate, follow-up observations can rule out sections of the parameter space of different astrophysical false positive scenarios. Since WASP-132 is already known to host a confirmed hot Jupiter, both speckle imaging and radial velocity follow-up observations were readily available for our analysis and could be included in our final validation of WASP-132 c.

\subsubsection{SOAR Speckle Imaging} \label{sssec:SOAR}

\begin{figure}
    \centering
    \includegraphics[width=0.5\textwidth]{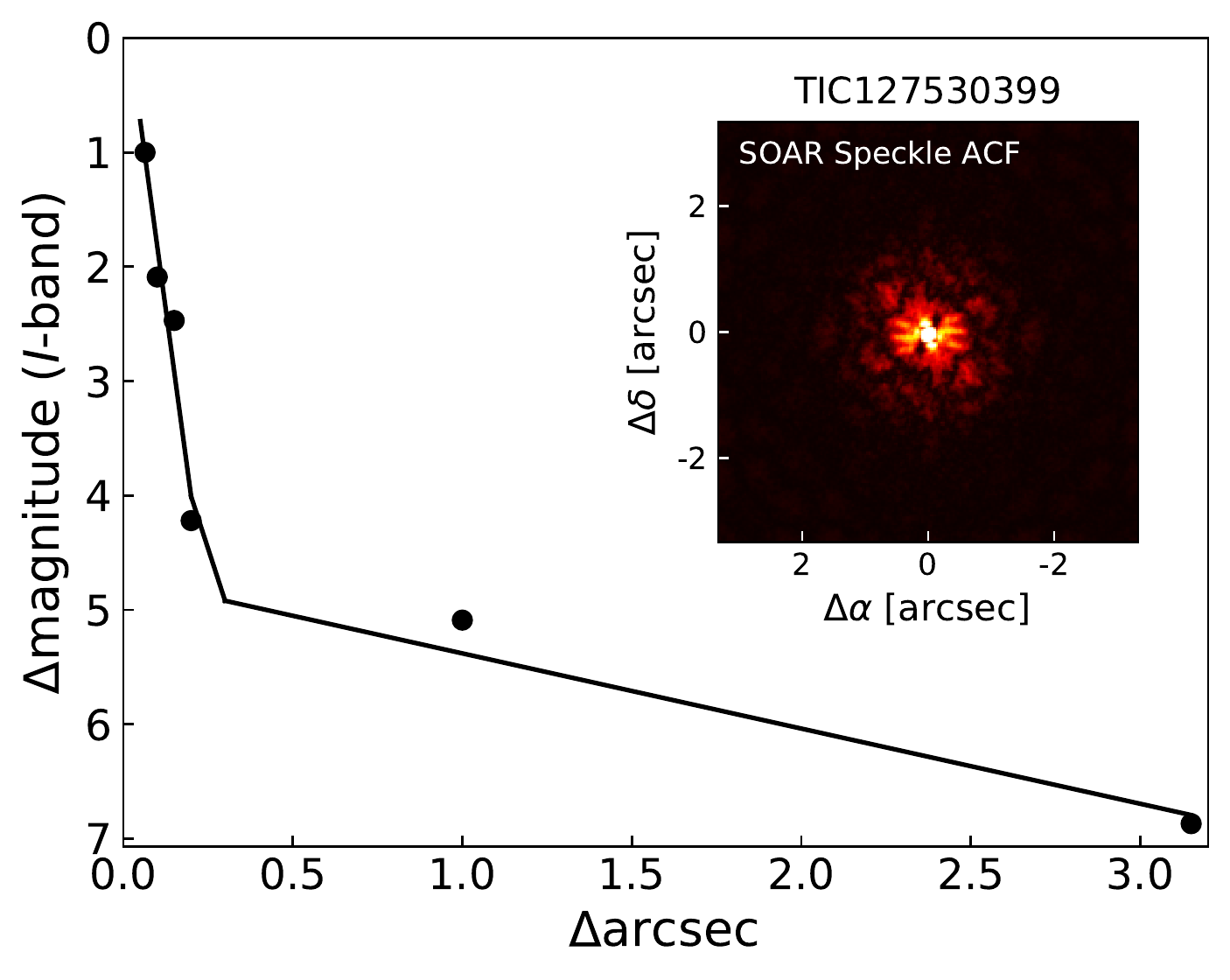}
    \caption{The 5-$\sigma$ detection sensitivity of the SOAR speckle imaging of WASP-132, with inset two-dimensional auto-correlation function reconstructed image of the field. The data indicate that there are no close-in companions within 3 arcsecond of WASP-132.}
    \label{fig:SOAR_curve}
\end{figure}

High-angular resolution imaging is needed to search for nearby sources that can contaminate the TESS photometry (resulting in an underestimated planetary radius) or be the source of astrophysical false positives, such as background eclipsing binaries. We searched for stellar companions to WASP-132 with speckle imaging using the HRCam instrument on the 4.1-m Southern Astrophysical Research (SOAR) telescope \citep{tokovinin2018ten} on 10 February 2020 UT, observing in Cousins I-band, a similar visible bandpass as TESS. This observation was sensitive to a 5.0-magnitude fainter star at an angular distance of 1 arcsec from the target. More details of the observation are available in \cite{2020AJ....159...19Z}. The 5$\sigma$ detection sensitivity and speckle auto-correlation functions from the observations are shown in Figure \ref{fig:SOAR_curve}. No nearby stars were detected within 3\arcsec of WASP-132 in the SOAR observations.

\subsubsection{CORALIE Radial Velocity Observations} \label{sssec:CORALIE}

In addition to SOAR speckle imaging, prior to the discovery of WASP-132 c there already existed radial velocity measurements of the host star WASP-132 which were used to confirm the planetary nature and measure the mass of the hot Jupiter WASP-132 b. In total, there were 36 radial velocity measurements taken across a time span of almost 2 years. All of these measurements were obtained with the CORALIE spectrograph which is an echelle spectrograph mounted on the 1.2-m Euler telescope in La Silla, Chile. These data are published in \cite{2017MNRAS.465.3693H}, which provides further information regarding the method used to reduce the radial velocity data.

According to the \texttt{forecaster} Python package \citep{chen2016probabilistic}, the projected mass of WASP-132 c should be 4.45 \mearth. \texttt{forecaster} assumed a gaseous envelope when estimating the mass of WASP-132 c, however the radius of the planet may place it in the super-Earth regime of planets. This would imply that the density of the planet is higher than a gaseous planet, meaning that the mass follows the relation $M_{\rm p}$ $\sim R_{\rm p}^{3.7}$. This would result in a mass of 9.74 \mearth. We made no distinction between these two scenarios in our analysis of CORALIE data since we imposed a very wide prior on the mass of WASP-132 c.

We first performed a joint fit of the time series \tess photometry with the CORALIE radial velocity measurements using \texttt{exoplanet}. This way, we were able to take advantage of the unique strengths of each data type in constraining the orbital and planet parameters as well as accounting for the noise and variability in the data. We used a model similar to that described in Section \ref{sec:modeling} but with the addition of planet mass as well as a quadratic trend and a jitter term for the radial velocities. The priors on the trend and jitter model components were Normal distributions with the quadratic trend distribution centered on 0 and the jitter term distribution centered on the standard deviation of the radial velocity data. We imposed a wide log-uniform prior on the mass from 0 to 8. This is more than wide enough to encompass the projected masses of both potential compositions for WASP-132 c as well as the 3$\sigma$ upper limit on the reported mass of WASP-132 b. CORALIE also underwent an upgrade partway through the dataset and this is accounted for by an offset between the pre- and post-upgrade portions of the data. We also fit for the trend in the data described in \cite{2017MNRAS.465.3693H}.

\begin{figure}
    \centering
    \includegraphics[width=0.45\textwidth]{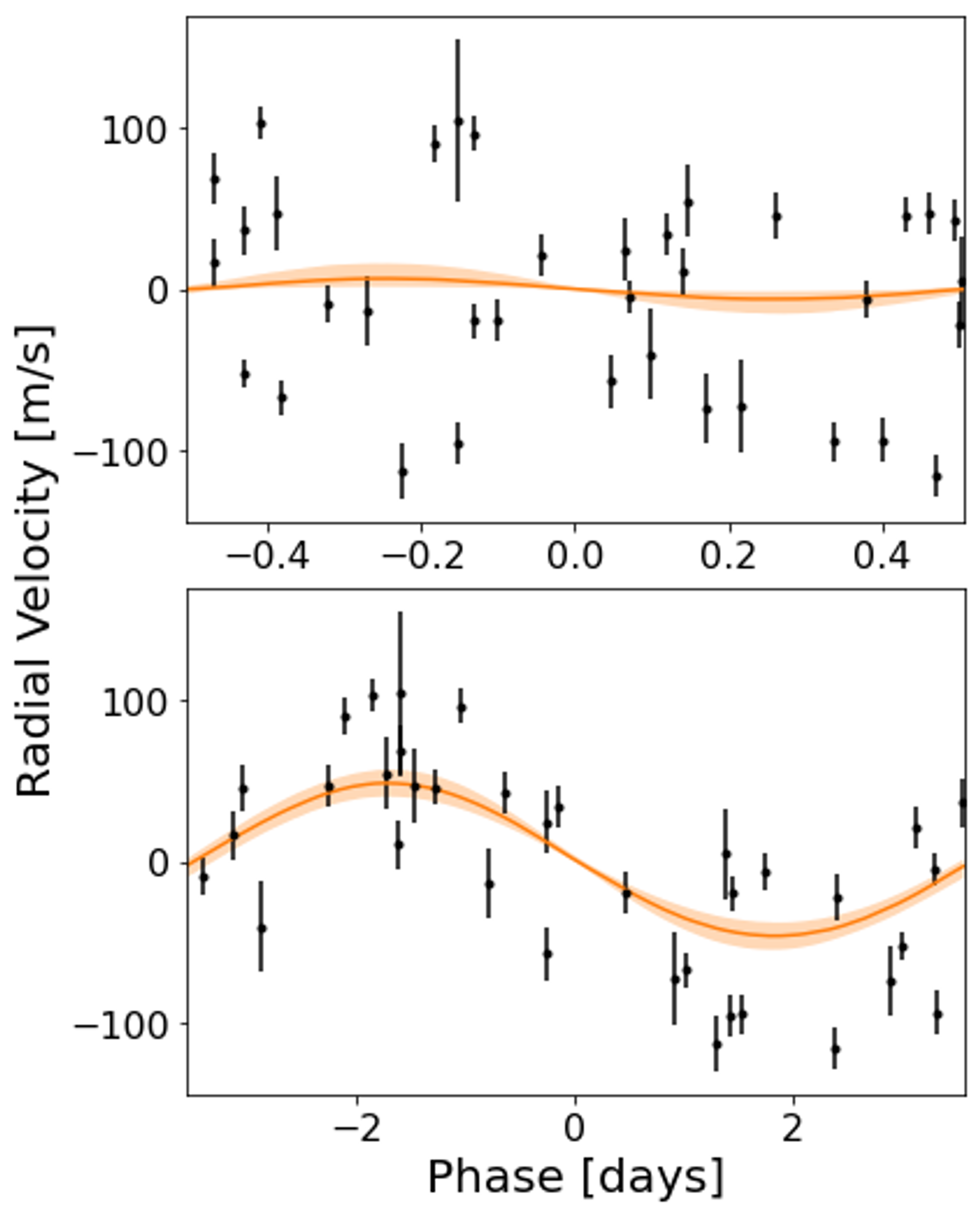}
    \caption{The CORALIE radial velocity measurements phase-folded to the best-fit periods of WASP-132 c (\textit{top}) and WASP-132 b (\textit{bottom}). The solid orange line represents the median radial velocity model and the shaded regions represent the 1$\sigma$ uncertainties in the model.}
    \label{fig:rv_phases}
\end{figure}

The posterior probability distribution for the mass of WASP-132 c was highest around zero with a wide spread of values, indicating a non-detection in the radial velocity. Therefore, we report a 3$\sigma$ upper limit from this distribution of 37.35 \mearth, corresponding to a radial velocity semi-amplitude of 28.23 ms$^{-1}$. All other planetary and orbital parameters modeled with the joint \tess + CORALIE data (e.g. period, mid-transit time, radius, etc.) are consistent with the values obtained from modeling the \tess data alone, described in Section \ref{sec:modeling}. The best fit models are plotted with the phase-folded radial velocity data for WASP-132 b and c in Figure \ref{fig:rv_phases}. We note that \cite{2017MNRAS.465.3693H} report that the CORALIE radial velocities show excess scatter, which may be due to magnetic activity. Therefore, the upper limit on the mass reported here may be inflated due to the high scatter.

As a check on the validity of the procedure, we also compared the measured planet mass for the confirmed hot Jupiter in the system WASP-132 b against its reported mass in \cite{2017MNRAS.465.3693H}. We measure the mass to be 121.89 $\pm$ 21.85 \mearth which is in agreement with the 130.31 $\pm$ 9.53 \mearth measured by \cite{2017MNRAS.465.3693H}. The uncertainty on our calculated mass of the hot Jupiter is higher than that reported by \cite{2017MNRAS.465.3693H} likely because \cite{2017MNRAS.465.3693H} also joint fit the CORALIE data with photometric data from both Wide Angle Search for Planets (WASP) and TRAnsiting Planets and PlanetesImals Small Telescope (TRAPPIST) observations that our analysis does not include as well as other ancillary RV observation data products.

As an independent check of the values obtained from \texttt{exoplanet}, we used the online Data and Analysis Center for Exoplanets\footnote{\url{https://dace.unige.ch}} (\textit{DACE}) platform to model the radial velocity data and investigate the significance of any signals around 1.01 d (the period of WASP-132 c). The Keplerian model initial conditions for the hot Jupiter were input based on the values on ExoFOP-TESS for WASP-132 b and a decrease of 60 $ms^{-1}$ over the course of the observations was added as noted in \cite{2017MNRAS.465.3693H}. When viewing the periodigram of the radial velocity data after adding in these components, there appear to be spikes in signal around 1 d, but none of which has an FAP $< 10\%$. This is not surprising, since many signals are aliased to 1 d and this is a common phenomenon in analyzing radial velocity data. The addition of another Keplerian orbit at the expected orbital period of WASP-132 c with the predicted semi-amplitude based on the mass predicted by \texttt{forecaster} results in a higher reduced $\chi^{2}$, indicating a worse fit.

According to Equation 14 in \cite{lovis2011radial}, the expected semi-amplitude of the radial velocity signal of WASP-132 c with 4.45 \mearth as predicted by \texttt{forecaster} is 3.36 ms$^{-1}$. If an Earth-like composition is assumed, the mass of 9.74 \mearth would result in a semi-amplitude of 7.44 ms$^{-1}$. CORALIE has been reported to have an individual measurement precision ranging between 3.5 and 6 ms$^{-1}$ \citep{rickman2019coralie}, putting the projected mass of WASP-132 c slightly below and slightly above the lower limit of CORALIE's detectable parameter space for gaseous and Earth-like compositions, respectively. Combined with the fact that the orbital period falls very close to the highly-aliased value of 1 d, it is logical that there was no significant detection of WASP-132 c in the CORALIE radial velocity data. Further radial velocity observations are necessary with a more precise instrument such as the High Accuracy Radial velocity Planet Searcher (HARPS) or the Carnegie Planet Finder Spectrograph (PFS) in order to obtain a mass measurement for the planet WASP-132 c.

\subsubsection{LCOGT 1\,m} \label{sssec:lco_photometry}

We observed TOI-822 c from the Las Cumbres Observatory Global Telescope \citep[LCOGT;][]{Brown:2013} 1.0\,m network node at the South Africa Astronomical Observatory on UTC 2022 March 5, 2022 March 8, and 2022 March 9 in Sloan $i'$ band. We used the {\tt TESS Transit Finder}, which is a customized version of the {\tt Tapir} software package \citep{Jensen:2013}, to schedule our transit observations. The 1\,m telescopes are equipped with $4096\times4096$ SINISTRO cameras having an image scale of $0\farcs389$ per pixel, resulting in a $26\arcmin\times26\arcmin$ field of view. The images were calibrated by the standard LCOGT {\tt BANZAI} pipeline \citep{McCully:2018}. Photometric data were extracted using {\tt AstroImageJ} \citep{Collins:2017}. The target star apertures exclude virtually all of the flux from the nearest Gaia EDR3 and TESS Input Catalog known neighbor (TIC 1051778874), which is $10\farcs0$ northwest of TOI-822. The light curve data are available on the ExoFOP-TESS website\footnote{https://exofop.ipac.caltech.edu/tess/target.php?id=127530399} \citep{ExoFOP2019-wy}.

The UTC 2022 March 5 observation used focused observations that were intended to saturate the target star for purposes of checking fainter stars within $2\farcm5$ of TOI-822 c for a potential nearby eclipsing binary (NEB) that could be the cause of the periodic detection in the TESS data. We rule out NEB signals, relative to the depth required in each neighboring star given its TIC version 8 delta TESS magnitude, by a factor of more than three times the RMS of the light curve scatter for all neighboring stars out to $50\arcsec$ from TOI-822. This is consistent with the SPOC pipeline's (lack of) centroid offset finding that the source of the TESS-detected signal is within $\sim10\arcsec$ of the target star location using data from TESS sectors 11 and 38. We also do not see any obvious NEB signals in stars out to $2\farcm5$, although some light curves suffer from blending from brighter nearby stars.

Although we expected the target star to be saturated in the UTC 2022 March 5 observation, it ultimately was strongly exposed, but not saturated. While the detrended light curve is consistent with a 40 min ($2.7\sigma$) late $\sim600$ ppm deep event on-target relative to the SPOC sectors 11 and 38 ephemeris, the detection is considered inconclusive due to limited post-transit baseline coverage and apparent systematics at the level of $\sim500$ ppm in the undetrended light curve.

The UTC 2022 March 8 and UTC 2022 March 9 observations were intentionally defocused to attempt to confirm the tentative $2.7\sigma$ late $\sim600$ ppm deep event in the UTC 2022 March 5 data. However, these data also suffered from $\sim 500$ ppm systematics in the undetrended light curves. With our best detrending efforts, the UTC 2022 March 8 would marginally suggest a roughly 100 min late $\sim 600-700$ ppm deep ingress at the end of the light curve. On the other hand, the UTC 2022 March 9 detrended light curve would suggest a roughly 20 min early $\sim700$ ppm deep event.

Given the level of systematics in the undetrended data and the inconsistent timing of the three tentative transit detections in the detrended data, we do not further consider the on-target results in the following analyses. Although we favor the interpretation that the extracted events are systematics driven, we cannot rule out the interpretation that some or all of the tentative detections are astrophysical and that the timing offsets are indicative of large TTVs in the system.

\subsection{Validation Using Software Tools} \label{ssec:software_validation}

While observational constraints can rule out portions of the parameter space where astrophysical false positives could exist, these observational limits are incomplete and do not rule out the entirety of the false positive parameter space. We are able to statistically analyze the remaining likelihood of false positive signals using publicly-available software. Using the available follow-up observations as constraints, we ran \vespa \citep{2012ApJ...761....6M, 2015ascl.soft03011M} and \triceratops \citep{2020ascl.soft02004G, 2021AJ....161...24G} to further establish the planetary nature of this signal.

\vespa compares the transit signal to a number of astrophysical false-positives including an unblended eclipsing binary (EB), a blended background EB (BEB), a hierarchical companion EB, and EB scenarios with a double-period. We ran \vespa using the \tess light curves detrended with the noise model described in Section \ref{sec:modeling} and phase-folded using the median values of the posteriors for the period and mid-transit time. The modeled transits of hot Jupiter WASP-132 b were also subtracted from the light curve. We included the maximum possible secondary depth of phase-folded features calculated by \texttt{DAVE} (Section \ref{sec:search_vetting}) and the SOAR I-band contrast curve (Section \ref{sssec:SOAR}) as observational constraints when calculating the false positive probability of the WASP-132 c signal. By default, \vespa simulates the background starfield within 1 square degree of the target, but we dictated that the maximum aperture radius interior to which the signal must be produced be $42^{\prime\prime}$, which is the radius of two \tess pixels and the maximum size of the aperture that the SPOC pipeline used to extract the light curves.

Using these inputs, we calculated the false positive probability (FPP) of the WASP-132 c signal to be 0.00193. The only false positive scenario with any remaining probability was the case of a blended background EB, however the probability for that scenario was $\ll$0.01 and is highly disfavored over the planet scenario. Since the overall FPP $\ll$0.01, this signal can be considered statistically validated by \vespa.

As an independent check, we ran the WASP-132 c signal through the statistical validation software \triceratops. This software is similar to \vespa in that it checks the signal against a set of scenarios that could produce transit-like signals. \triceratops was specifically designed for \tess observations and accounts for known nearby stars contained within the light curve extraction aperture as well as stars within $2.5^{\prime}$ of the target. This tool calculates both the FPP of the signal and the probability that the planet candidate is a false positive originating from a known nearby star, labeled the nearby FPP (NFPP). 

Since \triceratops is sensitive to the extraction aperture, we ran \triceratops using light curves extracted with two separate apertures for comparison and quality check. The first run used the apertures that the TESS SPOC pipeline used to extract the PDC$\_$SAP light curves, which were 12 and 10 pixels centered on the target position for Sectors 11 and 38, respectively. The second run used custom light curves extracted using reduced apertures that were smaller than those used by the TESS SPOC pipeline with 4 and 6 pixels centered on the target position for Sectors 11 and 38, respectively.

For each \triceratops run, we input the apertures used to extract the light curves, the I-band contrast curve from SOAR, and the median values for period and mid-transit time from the modeling performed in Section \ref{sec:modeling}. The light curves used were phase-folded at these period and mid-transit time values. For each set of apertures, we ran \triceratops twenty times and took the average FPP values. Using the apertures from the TESS SPOC pipeline, we obtained an FPP = 0.0126 $\pm$ 0.0003 for WASP-132 c. Using the reduced apertures, we obtained an FPP = 0.0107 $\pm$ 0.0004. Since the nearby field has been cleared of nearby and background eclipsing binary systems by the LCOGT observations (see Section \ref{sssec:lco_photometry}), SOAR observations (see Section \ref{sssec:SOAR}), and the the TESS SPOC centroid offset tests (which put the signal within 1 $\pm$ 3 arcsec of the target), we assume an NFPP value of 0.

According to \cite{2021AJ....161...24G}, for a signal to be statistically validated as a planet by \triceratops, it must have FPP $<$ 0.015. Our signal of 1.01 d meets the \triceratops FPP criterion for both sets of extraction apertures. For consistency, we also reran \vespa on the reduced aperture light curve obtained an FPP = $9.02 \times 10^{-5}$, still well below the \vespa statistical validation threshold. This difference in FPP between \vespa and \triceratops may stem from the fact that \triceratops uses the actual \tess aperture and background star population while \vespa simulates this with a extraction radius and TRILEGAL simulations. Therefore, given the \vespa validation, sufficiently low \triceratops FPP values, and strong constraints on nearby background stars in the field of view, we consider this signal to be statistically validated as a planet.

Furthermore, neither of these statistical validation packages account for the fact that this signal is a part of a multi-planet system. \cite{Lissauer2012, Lissauer2014} demonstrated that false positives are less likely in multi-planet systems and that the FPPs calculated without accounting for this fact should be treated as upper limits. An analysis of \tess multiplanet systems indicates that this ``multiplicity boost" may reduce these FPPs by a factor of $\sim20\times$ \citep{guerrero2021tess}, although the scarcity of inner companions to hot Jupiters suggests that the multiplicity boost should be smaller than this factor.

\section{Dynamical Stability} \label{sec:dynamics}

To probe the dynamical stability of the system, we simulated the system using \texttt{REBOUND} \citep{rebound}, a flexible N-body integrator written in both Python and C. We ran 30 iterations of the system, each time perturbing the mass, eccentricity, inclination, and periods of each of the planets within their 3$\sigma$ uncertainties to test stability at the extremes of the possible parameter space for both planets. Each iteration was integrated for simulation timescales of 100 Myr and a timestep of 0.2 d ($\sim$20$\%$ of WASP-132 c's best-fit orbital period) using the MERCURIUS integrator \citep{reboundmercurius}. 

\begin{figure*}
    \centering
    \includegraphics[width=0.99\textwidth]{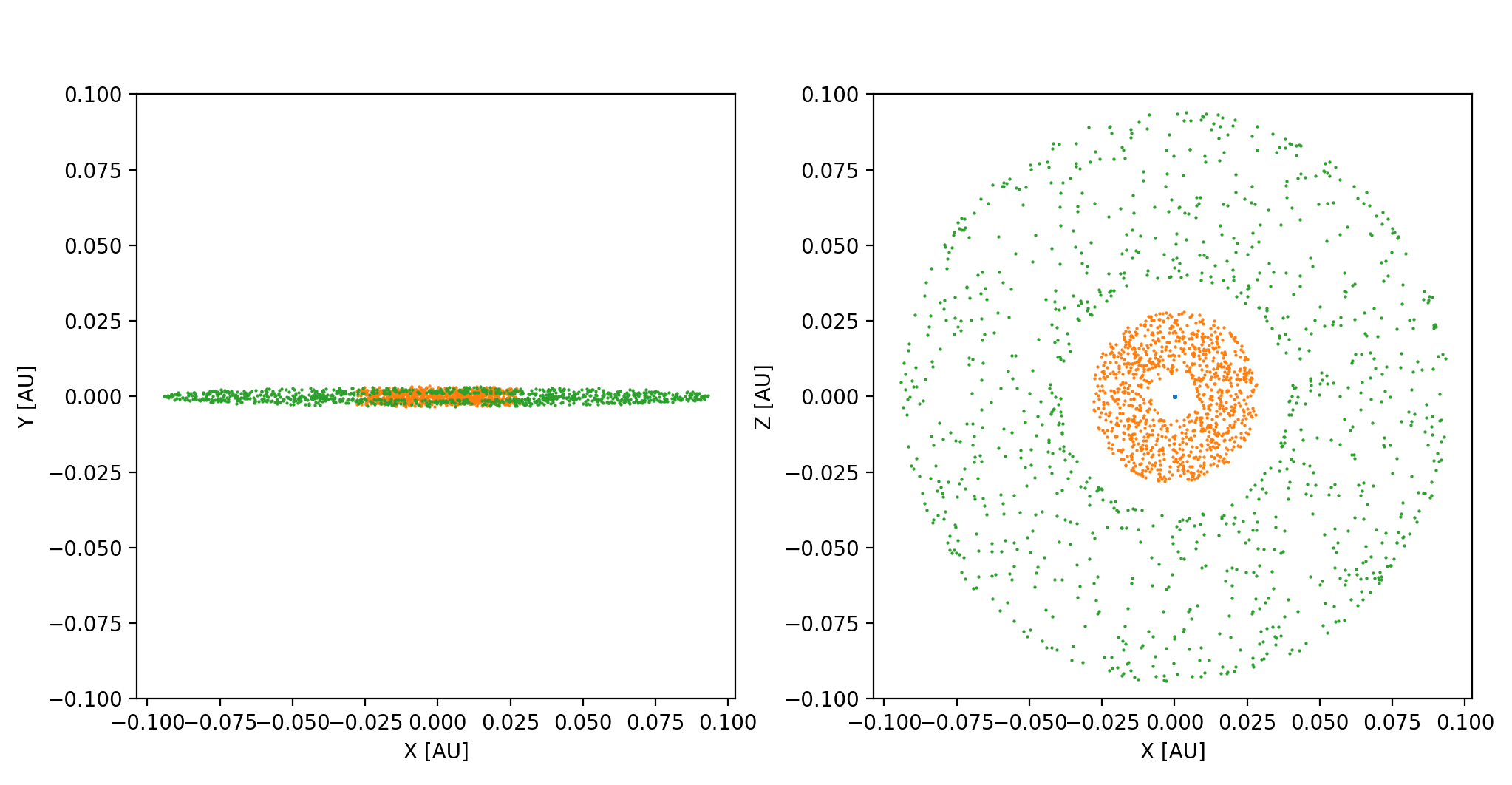}
    \caption{Position of WASP-132 b and c at 1000 evenly spaced points in time during the 1 Myr simulation with mass, eccentricity, and inclination initial values set to the 3$\sigma$ upper limits for both planets. WASP-132 b is denoted by the outer, green points and WASP-132 c is denoted by the inner, orange points. \textit{Left:} A side-on view of the system as it would be seen from Earth. \textit{Right:} A top-down view to illustrate the positions of both planets as they orbit over the course of 1 Myr. Note the gap between the maximum distance of WASP-132 c and the minimum distance of WASP-132 b from the host star.}
    \label{fig:dyn_sim}
\end{figure*}

The MERCURIUS integrator is a hybrid integrator that uses the symplectic Wisdom-Holman integrator WHFast when particles are far apart from each other and switches to the higher order integrator IAS15 during close encounters, which integrates with a smaller, adaptive timestep. We specified the minimum timestep for the IAS15 integrator as 0.02 d in order to speed up computation time.

We found that the planetary system is stable for all portions of the parameter space simulated on timescales of 100 Myr. None of the simulations exhibited a drastic change in semi-major axis outside of the normal oscillations due to gravitational interactions between the three bodies in the system (the two planets and the host star). Furthermore, no collisions were registered between any of the bodies over the course of the simulations. 

We also simulated the system using the 3$\sigma$ upper limits on mass, eccentricity, and inclination for both of the planets in order to maximize the likelihood of a close encounter. We integrated this system using the MERCURIUS integrator for 1 Myr at a timestep of 1.2 hours ($\sim$5$\%$ of WASP-132 c's best-fit orbital period). The positions of both planets at 1000 different evenly-spaced timesteps can be seen in Figure \ref{fig:dyn_sim}. We found that the system was stable at the extreme end of the parameter space of both planets as there were no close encounters between the planets and no significant change in the semi-major axis the two planets over the course of the simulation. This is illustrated by the gap between the innermost positions of WASP-132 b and outermost positions of WASP-132 c in Figure \ref{fig:dyn_sim}. Given the stability from both the simulations with random draws of parameters as well as the simulation with the planet parameter values most likely to cause a close encounter, the addition of WASP-132 c into the WASP-132 system does not appear to affect the long-term stability of the system.

\section{Discussion} \label{sec:discussion}

\begin{figure}
    \centering
    \includegraphics[width=0.5\textwidth]{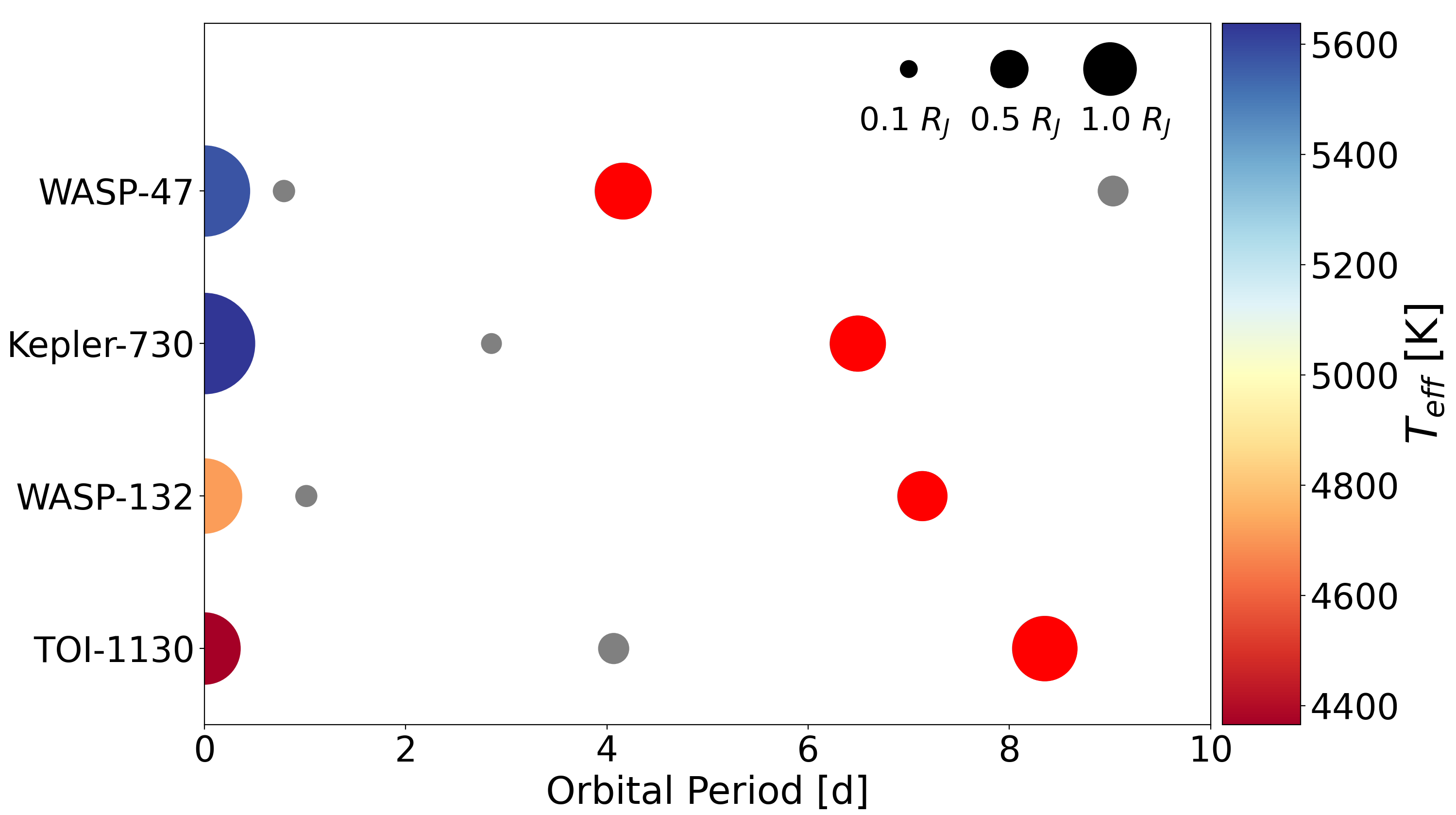}
    \caption{A schematic depiction of the four known systems with a hot Jupiter and at least one nearby companion planet (WASP-47, Kepler-730, TOI-1130, WASP-132). Orbital periods of the hot Jupiters increase from top to bottom. Sizes of circles for the planets are to scale with one another but not with distance from host star and host star radius. Likewise, circles for the host stars are to scale with one another but not with planets or orbital distance. Hot Jupiters are denoted by red circles, companion planets by gray circles, and the stellar hosts are colored based on their surface temperature.}
    \label{fig:hj_pictogram}
\end{figure}

The discovery and validation of WASP-132 c places the WASP-132 system among only a handful of systems with nearby companions to a hot Jupiter. Prior to WASP-132, only WASP-47 \citep{becker2015wasp}, Kepler-730 \citep{canas2019kepler}, and TOI-1130 \citep{huang2020tess} were known to harbor a hot Jupiter with nearby companions. Figure \ref{fig:hj_pictogram} illustrates all of the currently known systems containing hot Jupiters with nearby companions. These four multi-planet systems are still consistent with occurrence rate estimates provided by both \cite{huang2016warm}, \cite{zhu2021exoplanet}, and \cite{hord2021uniform} which calculate 1.1$\substack{+13.3 \\ -1.1}\%$, $\sim$2$\%$ ($<$9.7$\%$, 95$\%$ upper limit), and $7.3\substack{+15.2 \\ -7.3}\%$ of hot Jupiters to have nearby companions, respectively.

\subsection{Potential Formation Pathways}

The existence of nearby companions in four hot Jupiter systems suggests a dynamically cooler formation mechanism than the high-eccentricity migration that is invoked to explain a significant fraction of hot Jupiters. Although \tess appears to be improving our understanding of the architecture of hot Jupiter systems -- now with a second hot Jupiter companion discovery presented here -- hot Jupiters with companions are still a rarity, comprising only a small percentage of the nearly 500 hot Jupiters currently confirmed. Few hot Jupiter systems have yet to been searched for companions to this level of sensitivity, though. This scarcity of hot Jupiters with nearby companions suggests that high-eccentricity formation scenarios may dominate the observed hot Jupiter population. However, there is increasing evidence that there must be a subpopulation of hot Jupiters that do not form via high-eccentricity migration. This is evidenced not only by the existence of hot Jupiters with nearby companions as presented here, but also statistical simulations and analytical calculations show that high-eccentricity migration alone cannot reproduce the observed hot Jupiter population and require a mechanism such as disk migration \citep{dawson2014photoeccentric, anderson2016formation, munoz2016formation}. Although disk migration could explain these handful of unique systems, it is also possible that a super-Earth could have managed to exceed the threshold mass for runaway gas accretion, forming the hot Jupiter and leaving intact any planets orbiting interior \citep{lee2014make, batygin2016situ, huang2020tess}.

The subpopulation of hot Jupiters with companions that apparently did not undergo high-eccentricity migration may have more in common with warm Jupiters (10 $< P < $ 100 d) than their fellow hot Jupiters in terms of formation. In contrast to hot Jupiters, $\sim$50$\%$ of warm Jupiters have nearby companion planets \citep{huang2016warm}. Hot Jupiters with companions form a seemingly continuous period distribution with their warm Jupiter counterparts \citep{huang2020tess} and may represent the tail end of a wider population spanning across the physically-unmotivated 10 d dividing line.

Taking, for instance, a definition of hot Jupiter based on equilibrium temperature where $T_{\rm eq} > $1000 K defines a hot Jupiter (e.g. \citealt{miller2011heavy, thorngren2018bayesian}) leaves only WASP-47 b and Kepler-730 b classified as hot Jupiters with $T_{\rm eq}$ of 1259 and 1219 K, respectively, with TOI-1130 b and WASP-132 b classified in a slightly lower ``warm" range with $T_{\rm eq}$ of 637 and 763 K, respectively. This definition of hot Jupiters based on equilibrium temperature is equally arbitrary, however, as it still fails to provide a clear distinction between hot Jupiter subpopulations that formed via different formation mechanisms. Systems from a potential ``warm Jupiter tail", such as those discussed here, would inadvertently be included in this definition of hot Jupiters.

\begin{figure*}
    \centering
    \includegraphics[width=\textwidth]{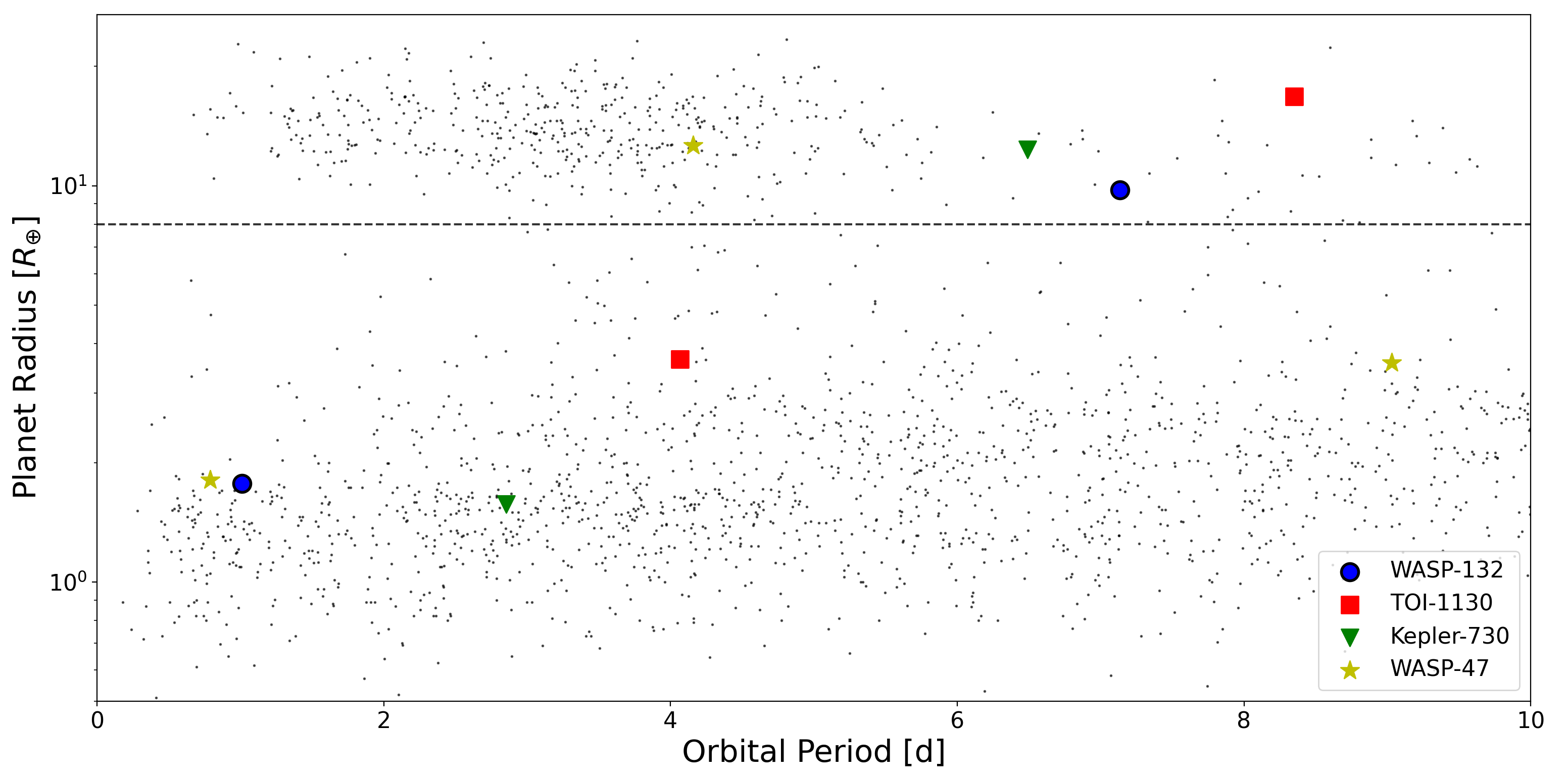}
    \caption{The period and radius of all planets in the four hot Jupiter systems with known companion planets (WASP-47, Kepler-730, TOI-1130, WASP-132) overlaid on the periods and radii of all planets contained on the NASA Exoplanet Archive. The larger, colored markers represent planets in one of these systems, with similar markers corresponding to planets in the same system. The markers for WASP-132 b and c are outlined in black. A dashed horizontal line has been included at 8\rearth to denote the division between hot Jupiters and smaller planets.}
    \label{fig:HJ_p-r}
\end{figure*}

\subsection{Investigating System Architectures}

Comparing these unique systems to the overall exoplanet population (see Figure \ref{fig:HJ_p-r}), it becomes apparent that three of the hot Jupiters with nearby companions have orbital periods that are longer than the typical hot Jupiter. The hot Jupiters with companions reside in a more sparsely-populated region of period-radius space farther from their host stars than most hot Jupiters. While this is notable, the sample of known hot Jupiters with nearby companions is currently too small to determine if there is a correlation between the presence of a companion and the hot Jupiter orbital or physical properties (or host star type). Furthermore, the cluster of hot Jupiters around an orbital period of 2-3 days is dominated by transit detections, whereas hot Jupiters are quite common on 4-5 day orbits when considering only radial velocity discoveries. Further analysis is required to quantify the extent to which the hot Juiters with close companions are outliers in the period-radius space.

What we know today is that the known hot Jupiter nearby companions predominantly orbit interior to the hot Jupiter, so the hot Jupiter must orbit at a certain distance from the host star so that there is enough room in the system for a stable orbit of a smaller planet interior to the hot Jupiter. There is the notable exception of the WASP-47 system, which also has one exterior companion. Whether additional exterior transiting companions to hot Jupiters exist remains to be seen, especially given the fact that all of the currently known close companions have been discovered via transits, which are biased towards planets on shorter orbital periods. With the probability of transit dropping with increasing orbital period, the discovery of additional exterior transiting companions is more challenging. Constraints have been placed on the existence of nearby, non-transiting planets through the investigation of transit timing variations (TTVs) using \textit{Kepler} data, with only WASP-148 c found \citep{steffen2012kepler, hebrard2020discovery}, supporting the idea that exterior companions are indeed perhaps intrinsically rarer than interior companions. Analysis of comprehensive transit timing searches using \tess data (such as that of \citealt{ivshina2022tess}) could provide updated constraints, however. 

Simulations performed by \cite{ogihara2013crowding, ogihara2014n} have provided a possible explanation for the dearth of external companions. They show that the existence of super-Earths exterior to a hot Jupiter would drive the hot Jupiter into the host star. Therefore, any hot Jupiters with external planets would have been driven into the host star prior to their discovery, thus leading to the ``crowding-out of giants by dwarfs", as \cite{ogihara2013crowding} refers to it.

In each of the previously known hot Jupiter systems with nearby companions, none of the planets are in orbital resonance with each other. The case is the same for WASP-132. A small fraction of \textit{Kepler} multiplanet systems are in resonances \citep{lissauer2011architecture}, whereas wide-orbiting giant planets frequently are \citep{winn2015occurrence}. Expanding the sample of hot Jupiters with nearby companions would allow for the comparison of the period ratio distribution of hot Jupiters with nearby companions to other known exoplanet populations. Similar distributions of period ratios would be expected from populations of systems that were assembled in a similar manner, and comparing the period ratio distribution of hot Jupiters with companions and the occurrence of resonances to a population such as super-Earth systems could elucidate possible common formation mechanisms.

\subsection{Prospects for Follow-Up}

These hot Jupiters with companions provide a unique opportunity to test if spin-orbit misalignment is a consequence of high-eccentricity migration as theorized. Based on the derived parameters of the host star and WASP-132 c, we estimate that the semi-amplitude of the Rossiter-McLaughlin effect for WASP-132 c is $\sim$0.5 m s$^{-1}$ (assuming a $v$ sin $i$ $\sim$0.9 km s$^{-1}$ as reported by \cite{2017MNRAS.465.3693H}). The estimated semi-amplitude of the hot Jupiter WASP-132 b is much larger at $\sim$17 m s$^{-1}$. Although the prospect of measuring the Rossiter-McLaughlin effect for WASP-132 c is quite low, that of WASP-132 b can easily be detected with current radial velocity precision and may inform our understanding of the orbital alignment with the host star's rotation. The WASP-132 system is otherwise amenable to additional follow-up observations as it is relatively bright at near-infrared wavelengths ($K_{\rm s}$ = 9.674).

Further radial velocity monitoring and a measured mass is necessary to confirm the planetary nature of this planet. Given that the expected semi-amplitude of WASP-132 c is between 3 and 6 ms$^{-1}$, an instrument more sensitive than CORALIE such as HARPS or PFS would be ideal to follow up on WASP-132 c to obtain a mass measurement and further characterize the system. Both of these instruments have 1 ms$^{-1}$ sensitivity or better, making a detection of WASP-132 c with either instrument possible.

\section{Summary} \label{sec:summary}

In this paper, we present the discovery and validation of a companion planet orbiting interior to hot Jupiter WASP-132 b. Our investigation and results are summarized here:

\begin{enumerate}
    \item A $\sim$1.01 d periodic signal with a false alarm probability $\ll$ 1$\%$ was detected in the \tess photometric data for WASP-132 (TOI-822) with both the SPOC pipeline and the Transit Least Squares search algorithm. Neither the TESS SPOC pipeline nor the vetting software \texttt{DAVE} found any immediate false positive indicators.
    \item We refined the system parameters, obtaining a host star of R$_{*}$ = $0.753\substack{+0.028 \\ -0.026}$ \rsun and M$_{*}$ = 0.782 $\pm$ 0.034 \msun and planets with periods and radii of 7.13 d and 10.05 $\pm$ 0.28 \rearth for WASP-132 b and 1.01 d and 1.85 $\pm$ 0.10 \rearth for WASP-132 c.
    \item An analysis of archival CORALIE radial velocity measurements did not yield a significant detection at the 1.01 d period, with a 3$\sigma$ upper limit of 37.35 \mearth on the mass of WASP-132 c.
    \item Using LCOGT ground-based follow-up photometry, we ruled out NEB signals as the potential source of the TESS detection in stars out to $50\arcsec$ from the WASP-132 c, and likely ruled out potential NEB signals in stars out to $2\farcm5$
    \item WASP-132 c is statistically validated as a planet with false positive probabilities (FPPs) of $9.02 \times 10^{-5}$ and 0.0107 using \vespa and \triceratops, respectively (see Section \ref{ssec:software_validation} for further discussion).
    \item The system is dynamically stable on timescales of 100 Myr for planetary and orbital parameters within 3$\sigma$ of the best-fit values.
    \item WASP-132 is the second system discovered by TESS (and one of only four systems in total) to contain a hot Jupiter with a nearby companion planet, suggesting that a mechanism other than high-eccentricity migration may play a significant role in the formation of hot Jupiters.
\end{enumerate}

The discovery of WASP-132 c demonstrates the ability of \tess to not only find new planets but also enhance our knowledge of those already known. As \tess continues its almost-all-sky survey, it will surely reveal additional systems similar to WASP-132 which will improve our understanding of the evolution of hot Jupiter systems. With an ever-expanding census of hot Jupiters with nearby companion planets, it may even be possible to identify sub-populations of these hot, giant planets. It is imperative to continue the search for this type of system architecture as a larger data set will allow us to solidify our understanding of how these rare systems form.

\acknowledgements
We thank J. Rowe and T. Barclay for the valuable suggestions for this study.

This paper includes data collected by the TESS mission, which are publicly available from the Mikulski Archive for Space Telescopes (MAST) and produced by the Science Processing Operations Center (SPOC) at NASA Ames Research Center. This research effort made use of systematic error-corrected (PDC-SAP) photometry. Funding for the TESS mission is provided by NASA's Science Mission directorate. 

Resources supporting this work were provided by the NASA High-End Computing (HEC) Program through the NASA Advanced Supercomputing (NAS) Division at Ames Research Center for the production of the SPOC data products.

This research has made use of the Exoplanet Followup Observation Program website, which is operated by the California Institute of Technology, under contract with the National Aeronautics and Space Administration under the Exoplanet Exploration Program.

This work has made use of data from the European Space Agency (ESA) mission {\it Gaia} (\url{https://www.cosmos.esa.int/gaia}), processed by the {\it Gaia} Data Processing and Analysis Consortium (DPAC, \url{https://www.cosmos.esa.int/web/gaia/dpac/consortium}). Funding for the DPAC has been provided by national institutions, in particular the institutions participating in the {\it Gaia} Multilateral Agreement.

This publication makes use of The Data $\&$ Analysis Center for Exoplanets (DACE), which is a facility based at the University of Geneva (CH) dedicated to extrasolar planets data visualisation, exchange and analysis. DACE is a platform of the Swiss National Centre of Competence in Research (NCCR) PlanetS, federating the Swiss expertise in Exoplanet research. The DACE platform is available at \url{https://dace.unige.ch}.

Simulations in this paper made use of the REBOUND N-body code \citep{rebound}. The simulations were integrated using the hybrid symplectic MERCURIUS integrator \citep{reboundmercurius}.

This work makes use of observations from the LCOGT network. Part of the LCOGT telescope time was granted by NOIRLab through the Mid-Scale Innovations Program (MSIP). MSIP is funded by NSF.

B.J.H. acknowledges support from the Future Investigators in NASA Earth and Space Science and Technology (FINESST) program - grant 80NSSC20K1551 - and support by NASA under award number 80GSFC21M0002. T.A.B. and M.L.S. acknowledge support from an appointment to the NASA Postdoctoral Program at the NASA Goddard Space Flight Center, administered by Universities Space Research Association and Oak Ridge Associated Universities under contract with NASA.

\facilities{CORALIE, \textit{Gaia}, LCOGT, SOAR, \tess}
\software{AstroImageJ \citep{Collins:2017}, astropy \citep{robitaille2013astropy, price2018astropy}, celerite \citep{foreman2017fast, 2018RNAAS...2...31F}, DACE, DAVE \citep{kostov2019b}, exoplanet \citep{exoplanet:exoplanet}, Forecaster \citep{chen2016probabilistic}, Jupyter \citep{kluyver2016jupyter}, Lightkurve \citep{2018ascl.soft12013L}, matplotlib \citep{hunter2007matplotlib}, NumPy \citep{van2011numpy}, Pandas \citep{pandas}, PyMC3 \citep{exoplanet:pymc3}, SciPy \citep{scipy}, STARRY \citep{exoplanet:luger18, exoplanet:agol19}, TAPIR \citep{Jensen:2013}, Theano \citep{exoplanet:theano}, TRICERATOPS \citep{2020ascl.soft02004G, 2021AJ....161...24G}, REBOUND \citep{rebound}, vespa \citep{2012ApJ...761....6M, 2015ascl.soft03011M}}

\bibliographystyle{aasjournal}
\bibliography{main}

\end{document}